\documentclass{article}

\usepackage[utf8]{inputenc}
\usepackage[T1]{fontenc}

\usepackage[margin=1.6in]{geometry}
\usepackage{needspace}

\usepackage{amsthm, amsmath, amssymb}
\usepackage{eucal}
\usepackage{complexity}

\usepackage{float}
\usepackage[table]{xcolor}
\usepackage{thmtools, thm-restate}
\usepackage{enumerate}
\usepackage{hyperref}
\usepackage{graphicx}
\usepackage{authblk} 
\usepackage{titlesec} 

\usepackage{array}

\usepackage{enumerate}
\usepackage{tabularx}
\usepackage{environ}
\usepackage{xspace}

\makeatletter
\newcommand{\problemtitle}[1]{\gdef\@problemtitle{#1}}%
\newcommand{\problemparameter}[1]{\gdef\@problemparameter{#1}}
\newcommand{\probleminput}[1]{\gdef\@probleminput{#1}}
\newcommand{\problemquestion}[1]{\gdef\@problemquestion{#1}}
\NewEnviron{problemgen}{
\problemtitle{}\problemparameter{}\probleminput{}\problemquestion{}
\BODY
\par\addvspace{.5\baselineskip}
\noindent
\begin{tabularx}{\textwidth}{@{\hspace{\parindent}} l X c}
    \multicolumn{2}{@{\hspace{\parindent}}l}{\@problemtitle} \\
    \textbf{Input:} & \@probleminput \\
    \textbf{Output:} & \@problemquestion
  \end{tabularx}
  \par\addvspace{.5\baselineskip}
}

\NewEnviron{problemdec}{
\problemtitle{}\probleminput{}\problemquestion{}
  \BODY
  \par\addvspace{.5\baselineskip}
  \noindent
  \begin{tabularx}{\textwidth}{@{\hspace{\parindent}} l X c}
    \multicolumn{2}{@{\hspace{\parindent}}l}{\@problemtitle} \\
    \textbf{Input:} & \@probleminput \\
    \textbf{Question:} & \@problemquestion
  \end{tabularx}
  \par\addvspace{.5\baselineskip}
}

\newcommand{\problemdot}{\(\cdot\allowbreak{}\)} 
\newcommand{\hornmssenum}{\textsc{Horn-MSS}\problemdot\textsc{Enum}}
\newcommand{\hornmssanother}{\textsc{Horn-MSS}\problemdot\textsc{AnotherSol}}

\newcommand{\bfa}{\mathbf{a}}
\newcommand{\bfb}{\mathbf{b}}

\newcommand{\intv}[1]{[#1]}

\renewcommand{\cS}{\mathcal{S}} 

\hypersetup{
  pdftitle = {A note on the intractability of enumerating the maximal satisfiable subformulas of a Horn expression},
  pdfauthor = {Oscar Defrain, Corentin Jazeron, and Kazuhisa Makino},
  colorlinks = true,
  linkcolor = black!30!red,
  citecolor = black!30!green
}

\declaretheorem[parent=section, name=Theorem]{theorem}

\makeatletter
\@ifundefined{newcounteralias}{}{%
  \renewcommand\thmt@autorefsetup{%
    \@xa\def\csname\thmt@envname autorefname\@xa\endcsname
      \@xa{\thmt@thmname}}%
}
\makeatother

\newtheorem*{lemma*}{Lemma}
\newtheorem{corollary}[theorem]{Corollary}

\newtheorem{claim}[theorem]{\normalfont Claim}

\newenvironment{claimproof}{\begin{proof}[Proof of
    claim]}{\end{proof}}

\renewenvironment{abstract}
{\small\vspace{-1em}
\begin{center}
\bfseries\abstractname\vspace{-.5em}\vspace{0pt}
\end{center}
\list{}{
\setlength{\leftmargin}{0.6in}%
\setlength{\rightmargin}{\leftmargin}}%
\item\relax}
{\endlist}

\title{A note on the intractability of enumerating\\ the maximal satisfiable subformulas\\ of a Horn expression}

\author[1]{Oscar Defrain}
\author[1,2]{Corentin Jazeron}
\author[3]{Kazuhisa Makino}

\affil[1]{Aix-Marseille Université, CNRS, LIS, Marseille, France.}
\affil[2]{École Normale Supérieure Paris-Saclay, Gif-sur-Yvette, France.}
\affil[2]{Kyoto University, RIMS, Japan.}

\date{November 10, 2026} 

\begin{document}

\maketitle

\begin{abstract}
We show that, given a Horn expression and a family of inclusion-wise maximal satisfiable subsets of its clauses, deciding whether there exists another maximal satisfiable subset is \NP-complete.
As a consequence, unless $\P=\NP$, there is no output-polynomial time algorithm enumerating the maximal satisfiable subformulas of a Horn expression.
\end{abstract}

\section{Introduction}

A maximal satisfiable subformula (MSS) of a CNF formula $\phi$ is an inclusion-wise maximal subset of its clauses whose conjunction is satisfiable.
The MSS enumeration problem asks to list all such subsets without repetition.
Since their number may be exponential in the size of the formula at hand, a natural objective is to obtain \emph{output-polynomial time} algorithms, i.e., algorithms whose running times are polynomial in the combined size of the input and the output.
More desirable enumeration algorithms are \emph{incremental-polynomial time} algorithms that produce the $i$th solution in polynomial time in the input size plus $i$, or \emph{polynomial-delay} algorithms whose running times before the first output, between consecutive outputs, and after the last output, are bounded by a polynomial in the input size only;
see~\cite{johnson1988generating,strozecki2019survey} for more details on enumeration problems and complexities.

The problem of enumerating MSS can also be expressed as the enumeration of \emph{maximal signatures}, where a signature is a binary sequence that records the clauses satisfied by a truth assignment, and maximality is taken bit-wise.
B\'erczi et al.~\cite{berczi2021generating} showed that all signatures of a $k$-CNF can be enumerated in incremental-polynomial time for fixed $k$, and with polynomial delay for tractable CNFs.
A family of CNFs is said to be \emph{tractable} if for any CNF $\phi$ in this family, the satisfiability of any subformula of $\phi$ can be decided in polynomial time even after fixing any subset of the variables at arbitrary values.
In the same paper, the authors show that bit-wise minimal signatures of an arbitrary CNF can be listed with polynomial delay, by reducing to maximal independent sets enumeration in graphs, while the maximal ones cannot be enumerated in output polynomial-time unless $\P=\NP$, even for $3$-CNFs.
The latter intractability result follows from the fact that a formula $\phi$ admits a unique maximal signature if and only if the sequence $(1,\dots,1)$ is a signature, i.e., if and only if $\phi$ is satisfiable, and thus relies on the intractability of the satisfiability problem.

The latter observation raises the question of whether the MSS can be enumerated in output-polynomial time for tractable families of CNF formulas.
Natural cases include the Horn and $2$-CNF formulas.
Another natural setting, arising from Schaefer's dichotomy theorem~\cite{schaefer1978complexity}, is that of XOR formulas, whose clauses use the exclusive-or connective and describe parity constraints.
Creignou et al.~\cite{creignou2025signatures} showed that maximal signatures of XOR formulas can be enumerated in incremental-polynomial time, and with polynomial delay if the XOR clauses have size at most 2.
In this note, we prove that there is no output-polynomial time algorithm to enumerate maximal satisfiable subformulas of a Horn expression unless $\P=\NP$.
Thus, the polynomial-time decidability of the satisfiability problem, even after a partial assignment has been fixed, does not suffice for output-polynomial time MSS enumeration.
This answers a question that was raised implicitly in \cite{berczi2021generating}, explicitly in \cite{creignou2025signatures}, and at several editions of the Workshop on Enumeration Problems and Applications (WEPA).

\section{Preliminaries}

\emph{Horn clauses} are non-empty disjunctions of literals containing at most one positive literal. 
Thus, they can consist of a unit clause containing a positive literal $(x)$, a fully negative clause of unbounded size $(\overline{x_1}\vee \dots\vee \overline{x_k})$, or a clause of unbounded size with exactly one positive literal $(\overline{x_1}\vee \dots\vee \overline{x_k}\vee y)$.
Clauses of the last type are usually seen as implications and are also denoted as $\{x_1,\dots,x_k\}\to y$.
A \emph{Horn expression} (also called \emph{Horn CNF}, or \emph{Horn formula}) is a conjunction of Horn clauses.

If $\phi$ is a Horn expression with clauses $C_1,\dots,C_m$, a maximal satisfiable subformula of $\phi$ is an inclusion-wise maximal set $S\subseteq \{C_1,\dots,C_m\}$ whose conjunction $\bigwedge S$ defines a satisfiable Horn formula.

In this note, we consider the following two problems.

\begin{problemgen}
  \problemtitle{Maximal Satisfiable Subformulas of a Horn expression (\hornmssenum)}
  \probleminput{a Horn CNF $\phi$.}
  \problemquestion{the maximal satisfiable subformulas of $\phi$.}
\end{problemgen}

\begin{problemdec}
  \problemtitle{Another solution for \hornmssenum{} (\hornmssanother)}
  \probleminput{a Horn CNF $\phi$ and a set $\cS$ of its maximal satisfiable subformulas.}
  \problemquestion{is there a maximal satisfiable subformula of $\phi$ that is not in $\cS$.}
\end{problemdec}

Note that given a Horn formula $\phi$, and since Horn satisfiability is polynomial-time decidable \cite{schaefer1978complexity}, one can greedily compute in polynomial time one of its maximal satisfiable subformulas, by iteratively adding a clause while the obtained set is satisfiable.
Moreover, by branching on the clauses that are not part of a solution, one can see that computing constantly many distinct solutions is also tractable.
In this note, we however show that computing a next solution, when the number of already obtained solutions is unbounded, is intractable.

\section{The proof}

\begin{theorem}\label{thm:main}
    \hornmssanother{} is \NP-complete.
\end{theorem}

\begin{proof}
    The fact that \hornmssanother{} is in \NP{} follows from the fact that a maximal satisfiable subformula of $\phi$ that is not in $\cS$ defines a polynomial-size certificate whose validity can be checked in polynomial time.
    
    We prove the \NP-hardness by providing a polynomial-time reduction from $3$-SAT.
    Let $\psi=\bigwedge_{j=1}^m C_j$ be an $n$-variable $m$-clause $3$-CNF with $n,m\geq 2$, and for every $j\in\intv{m}$, let us fix an arbitrary order on the literals of $C_j$ and denote them by $\ell_{j,1}, \ell_{j,2}, \ell_{j,3}$.
    Let us make the additional assumption that the variables of $\psi$ appear both positively and negatively. 
    Note that this assumption can be made without loss of generality, as variables appearing only positively can be set to true, and those appearing only negatively can be set to false, so that the clauses in which they belong are satisfied, and with no impact on the satisfiability of the remaining part of the formula.
    
    We construct an instance of \hornmssanother{} as follows.
    For every $j\in\intv{m}$, let $V_j:=\{v_{j,1}, v_{j,2}, v_{j,3}\}$ where $v_{j,i}$ is a variable that represents the occurrence of literal $\ell_{j,i}$ in $C_j$.
    Let us point that to distinct $i$'s and $j$'s we associate distinct variables $v_{j,i}$ even if the corresponding literals coincide.
    Let $V:=\bigcup_{j=1}^m V_j$, then $|V|=3m$.
    Given $j\in\intv{m}$ and $i\in\intv{3}$, let $\smash{L_{j,i}:=\{v_{j',i'} : \ell_{j',i'} = \overline{\ell_{j,i}}\}}$ denote the set of variables in $V$ that represent an occurrence of the negation of the literal represented by $v_{j,i}$.
    We define three families of Horn clauses as follows:
    \begin{align*}
        A&:= \Big\{(v_{j,i}) : j\in\intv{m},\ i\in\intv{3}\Big\},\\
        B&:= \Big\{\bigvee_{\substack{j\in\intv{m}\\ i\in\intv{3}}} \overline{v_{j,i}}\Big\},\\
        C&:= \Big\{L_{j,i} \cup V_{j-1} \to v_{j,i} : j\in\intv{m},\ i\in\intv{3}\Big\}.
    \end{align*}
    where by $V_{j-1}$ for ${j=1}$ is meant $V_m$, i.e., the set $C$ defines a circular set of implications. 
    Note that $|A|=|C|=3m$ and $|B|=1$.
    Let $\phi$ be the Horn expression defined on variable set $V$ and consisting of the set of clauses $A\cup B\cup C$.
    Then $\phi$ is a $3m$-variable $6m+1$-clause Horn expression.

    Let us further define:
    \begin{align*}
        \cS_1&:= \big\{A\cup C\big\},\\
        \cS_2&:= \big\{(A\cup B\cup C) 
        \setminus \{(v_{j,i}),\ L_{j,i}\cup V_{j-1}\to v_{j,i}\} : j\in [m],\ i\in [3]\big\},\\
        \cS_3&:= \big\{(A\cup B\cup C) 
        \setminus \{(v_{j,i}),(v_{j',i'})\} : j\in [m],\ i\in [3]\ \text{and}\ v_{j',i'}\in L_{j,i}\big\}.
    \end{align*}
    and put $\cS:=\cS_1\cup \cS_2\cup \cS_3$.
    Note that $|\cS|=O(n+m^2)$. 
    This concludes the construction of the instance of \hornmssanother{}, whose input validity is argued in the claim below.

    Let $\cS^*$ be the set of maximal satisfiable subformulas of $\phi$.

    \begin{claim}\label{claim:valid-instance}
        $\cS_1\cup \cS_2\cup \cS_3\subseteq \cS^*$.
    \end{claim}

    \begin{claimproof}
        We first consider $\cS_1$, which contains the single set $A\cup C$.
        Note that this set defines a satisfiable subformula, as all its clauses involve at least one positive literal.
        To see that it is maximal, note that $\bigwedge A$ and $B$ are contradictory.
        This proves $\cS_1\subseteq \cS^*$.

        We now consider $S\in \cS_2$.
        Then $S=(A\setminus \{(v_{j,i})\}) \cup B \cup (C\setminus \{L_{j,i}\cup V_{j-1}\to v_{j,i}\})$ for some variable $v_{j,i}$.
        We argue that $S$ is satisfiable by describing an assignment.
        Let $\bfa$ be the assignment that maps all variables of $V$ to $1$, except for $v_{j,i}$ which is mapped to $0$.
        Clearly, the clauses in $A\setminus \{(v_{j,i})\}$ are satisfied.
        The clause in $B$ is also satisfied as it contains $\overline{v_{j,i}}$.
        As of the clauses in $C$, they consist in implications having variables $v_{j',i'}$ for conclusion. 
        Hence all such clauses, except possibly for $L_{j,i}\cup V_{j-1}\to v_{j,i}$, are satisfied.
        Thus $\bfa$ satisfies $S$.
        Now, note that since $A\setminus \{(v_{j,i})\}$ consists of unit clauses, any assignment $\bfa'$ satisfying $S$ must actually map the variables of $V\setminus \{v_{j,i}\}$ to $1$.
        Moreover, since $B\subseteq S$, any such assignment $\bfa'$ must map the remaining variable $v_{j,i}$ to $0$.
        We conclude that there is a unique assignment $\bfa'=\bfa$ satisfying $S$, from which we easily see that neither $(v_{j,i})$ nor $L_{j,i}\cup V_{j-1}\to v_{j,i}$ can be added to $S$ while keeping its satisfiability.
        We conclude that $S$ is maximal, proving $\cS_2\subseteq \cS^*$.

        It remains to consider a set $S\in \cS_3$.
        Then $S=(A\setminus \{(v_{j,i}),(v_{j',i'})\})\cup B\cup C$ for some pair of variables $v_{j,i}, v_{j',i'}$ such that $v_{j',i'}\in L_{j,i}$.
        We argue that $S$ is satisfiable by describing an assignment.
        Let $\bfa$ be the assignment that puts every variable to $1$ except for the variables $v_{j,i}, v_{j',i'}$ that are mapped to $0$.
        Note that all the clauses in $A\setminus \{(v_{j,i}),(v_{j',i'})\}$ are satisfied by this assignment.
        (We note that this set may contain the clauses corresponding to other variables $v_{j'',i''}$ of $L_{j,i}$ distinct from $v_{j',i'}$ but this is not a problem.)
        Also, the clause in $B$ is satisfied.
        As of the clauses in $C$ they consist in implications having variables $v_{j'',i''}$ for conclusion. 
        All such implications with $v_{j'',i''}$ distinct from $v_{j,i}$ and $v_{j',i'}$ are thus satisfied.
        As for those having $v_{j,i}$ (resp.~$v_{j',i'}$) for conclusion, note that they also have $v_{j',i'}$ (resp.~$v_{j,i}$) in their premise, as part of $L_{j,i}$ (resp.~$L_{j',i'}$).
        (In fact, $v_{j',i'}$ is the only variable of such a premise to be set to $0$ by $\bfa$.)
        Hence they are also satisfied.
        Consequently, all implications in $C$ are satisfied, and we conclude that $\bfa$ satisfies $S$.
        Now, note that since $A\setminus \{(v_{j,i}),(v_{j',i'})\}$ consists of unit clauses, any assignment $\bfa'$ satisfying $S$ must map the variables of $V\setminus \{v_{j,i},v_{j',i'}\}$ to $1$.
        Suppose toward a contradiction that one such assignment $\bfa'$ satisfying $S$ could map one of $v_{j,i},v_{j',i'}$, say $v_{j',i'}$, to $1$.
        Then, since $\bfa'$ satisfies $S$ and $L_{j,i}\cup V_{j-1}\to v_{j,i} \in S$, and since $L_{j,i}\cup V_{j-1}\subseteq V\setminus \{v_{j,i}\}\subseteq \bfa'^{-1}(1)$, we would derive that $\bfa'$ maps $v_{j,i}$ to $1$ to satisfy the implication, and consequently that $\bfa'$ maps all variables of $V$ to $1$.
        This contradicts the fact that the clause of $B$ belongs to $S$.
        Hence there is a unique assignment $\bfa'=\bfa$ satisfying $S$, from which we derive that $(v_{j,i}),(v_{j',i'})$ cannot be added to $S$ while making it satisfiable.
        So $S$ is maximal, proving $\cS_3\subseteq \cS^*$ and concluding the proof.
    \end{claimproof}

    \begin{claim}\label{claim:iff}
        There exists $S\in \cS^* \setminus (\cS_1\cup \cS_2\cup \cS_3)$ if and only if $\psi$ is satisfiable.
    \end{claim}

    \begin{claimproof}
        We first prove the left-to-right implication.
        Let us assume that there exists $S\in \cS^* \setminus (\cS_1\cup \cS_2\cup \cS_3)$.
        By maximality of $S$ and because $S\not\in \cS_1$, we have that $B\subseteq S$.
        
        Let us show that $C\subseteq S$.
        Suppose toward a contradiction that this is not the case.
        Then there is a clause $L_{j,i}\cup V_{j-1}\to v_{j,i}$ of $C$ that is not in $S$.
        By maximality of $S$, because that implication cannot be added while keeping $S$ satisfiable, it must be that $v_{j,i}$ equals $0$ in any truth assignment $\bfa$ satisfying $S$.
        Hence the clause $(v_{j,i})\in A$ does not belong to $S$.
        Since neither $(v_{j,i})$ nor $L_{j,i}\cup V_{j-1}\to v_{j,i}$ belong to $S$, we conclude that $S$ is a subset of a solution in $\cS_2$, a contradiction.
        We conclude that $C\subseteq S$ as desired.

        So far, we proved that $S$ contains $B\cup C$ as a subset.
        We now claim that the remaining set of clauses $I:=A\setminus S$ defines a partial satisfying truth assignment $\bfb$ of $\psi$ that maps each of its variables $x$ to $1$ if there exists a clause $(v_{j,i})\in I$ such that $\ell_{j,i}=x$, or to $0$ otherwise (including the case $(v_{j,i})\in I$ and $\ell_{j,i}=\overline{x}$).
        
        Note first that $I$ does not contain two clauses $(v_{j,i}),(v_{j',i'})$ that represent contradicting literals, i.e., such that $v_{j',i'}\in L_{j,i}$, as otherwise $S$ is a subset of a solution in $\cS_3$. 
        Thus, if $(v_{j,i})$ is part of $I$, then either $\ell_{j,i}=x$ and $x$ is mapped to $1$ by $\bfb$, or $\ell_{j,i}=\overline{x}$ and $x$ is mapped to $0$ by $\bfb$.
        So $I$ defines a partial truth assignment of $\psi$.
        We also deduce that if $I$ contains a clause $(v_{j,i})$, then $\{(v_{j',i'}) : v_{j',i'}\in L_{j,i}\}\subseteq S$.
        
        Moreover, note that whenever $I$ contains a clause $(v_{j,i})$, since that clause is not part of $S$ and $S$ is maximal, any assignment $\bfa$ satisfying $S$ must map its corresponding variable $v_{j,i}$ to $0$. 
        Moreover, since $L_{j,i}\cup V_{j-1}\to v_{j,i}\in S$ and $\bfa$ satisfies $S$, it must be that at least one element of the premise $L_{j,i}\cup V_{j-1}$ is mapped to $0$ by $\bfa$. 
        Since $\{(v_{j',i'}) : v_{j',i'}\in L_{j,i}\}\subseteq S$, and since these clauses are unit clauses, $\bfa$ must maps these variables $v_{j',i'}$ to $1$. 
        We derive that the elements of the premise mapped to $0$ by $\bfa$ belong to $V_{j-1}$.
        Hence their corresponding clause in $A$ belong to $I$.
        Consequently, under the assumption that $I$ contains a clause $(v_{j,i})$, it must contain the clauses of $A$ corresponding to at least one variable in $V_{j-1}$, at least one variable in $V_{j-2}$, and so on.
        By the cyclicity of the implications in $C$, $I$ must actually contain the clause of at least one variable in each of $V_1,\dots,V_m$, showing that $\bfb$ is a satisfying truth assignment of $\psi$.

        We conclude observing that $I$ must contain at least one such clause of $A$ since $S$ contains $B$, and $\bigwedge A$ and $B$ are contradictory.

        \medskip

        We now turn to the right-to-left implication.
        Let us assume that there exists a satisfying truth assignment $\bfb$ for $\psi$.
        We set $I:=\{(v_{j,i}) : \ell_{j,i}\ \text{is satisfied by}\ \bfb\}$, and $S=(A\cup B\cup C)\setminus I$.
        Then $B\cup C\subseteq S$.
        Consider the truth assignment $\bfa$ which maps every variable $v_{j,i}\in V$ to $0$ if $(v_{j,i})\in I$, and to $1$ otherwise, i.e., if $(v_{j,i})\in S$.
        Let us first show that $\bfa$ satisfies $S$.

        By construction, the clauses in $S\cap A$ are satisfied by $\bfa$.
        As by assumption $\psi$ contains at least one literal and its negation, the constructed assignment $\bfa$ maps at least one variable to $0$, and so it satisfies $B$ as well. 
        So $\bfa$ satisfies $S\cap (A\cup B)$.
        Since $\bfb$ is a satisfying truth assignment of $\psi$, $I$ contains a clause $(v)$ for at least one variable $v$ in each of the sets $V_1,\dots,V_m$.
        Hence, $\bfa$ maps to $0$ at least one variable $v$ in each of the sets $V_1,\dots,V_m$.
        Hence, for every implication $L_{j,i}\cup V_{j-1}\to v_{j,i}$, an element of the premise is set to $0$ by $\bfa$.
        Consequently $\bfa$ satisfies all the implications in $C$, and so $\bfa$ satisfies $S$.

        Now, observe that since $S$ contains $B$, it is not a subset of a solution in $\cS_1$.
        Since it contains $C$, it is not a subset of a solution in $\cS_2$. 
        Finally, since it contains one of every pair of clauses $(v_{j,i}),(v_{j',i'})$ such that $v_{j',i'}\in L_{j,i}$, it is not a subset of a solution in $\cS_3$. 
        Consequently, any greedy completion of $S$ leads to a maximal satisfiable subformula of $\phi$ which is not in $\cS_1\cup \cS_2\cup \cS_3$.
        This completes the proof.
    \end{claimproof}

    Note that, given an $n$-variable $m$-clause $3$-CNF $\psi$, one may construct in polynomial time in $n+m$ a Horn expression $\phi$ and the sets $\cS_1, \cS_2$ and $\cS_3$ as defined above, all of which of polynomial size in $n+m$.
    Claim~\ref{claim:valid-instance} shows that the obtained sets indeed define a valid instance of \hornmssanother{}.
    Claim~\ref{claim:iff} shows that the existence of an algorithm solving \hornmssanother{} in polynomial time provides one deciding the satisfiability of $\psi$.
    This concludes the proof.
\end{proof}

\needspace{2cm}

We derive the following using the folklore observation that the existence of an output-polynomial time algorithm for an enumeration problem yields an (input) polynomial one for deciding whether there exists another solution; see~\cite{strozecki2019survey}.

\begin{corollary}\label{cor:main}
    There is no output-polynomial time algorithm for \hornmssenum{} unless $\P=\NP$.
\end{corollary}

We finally note that as $3$-SAT remains \NP-complete when each variable occurs in at most four clauses \cite{tovey1984simplified}, Theorem~\ref{thm:main} and Corollary \ref{cor:main} even hold for Horn expressions consisting of clauses of size at most $7$ plus a single clause of unbounded size.

\paragraph{Acknowledgments.}
The authors would like to thank Alessio Conte, Alberto L'Epi\-scopo, Kazuhiro Kurita, Arnaud Mary, Giulia Punzi, Takeaki Uno, Simon Vilmin, and Kunihiro Wasa for discussions on this problem and early attempts to obtain output-polynomial time algorithms.

\bibliographystyle{alpha}
\bibliography{main}

@article{berczi2021generating,
  title={Generating clause sequences of a {CNF} formula},
  author={B{\'e}rczi, Krist{\'o}f and Boros, Endre and {\v{C}}epek, Ond{\v{r}}ej and Elbassioni, Khaled and Ku{\v{c}}era, Petr and Makino, Kazuhisa},
  journal={Theoretical computer science},
  volume={856},
  pages={68--74},
  year={2021},
  publisher={Elsevier}
}

@InProceedings{creignou2025signatures,
  author =  {Creignou, Nadia and Defrain, Oscar and Olive, Fr\'{e}d\'{e}ric and Vilmin, Simon},
  title = {{On the Enumeration of Signatures of XOR-CNF’s}},
  booktitle = {19th International Symposium on Algorithms and Data Structures (WADS 2025)},
  pages = {19:1--19:14},
  series =  {Leibniz International Proceedings in Informatics (LIPIcs)},
  ISBN =  {978-3-95977-398-0},
  ISSN =  {1868-8969},
  year =  {2025},
  volume =  {349},
  editor =  {Morin, Pat and Oh, Eunjin},
  publisher = {Schloss Dagstuhl -- Leibniz-Zentrum f{\"u}r Informatik},
  address = {Dagstuhl, Germany},
  URL =   {https://drops.dagstuhl.de/entities/document/10.4230/LIPIcs.WADS.2025.19},
  URN =   {urn:nbn:de:0030-drops-242508},
  doi =   {10.4230/LIPIcs.WADS.2025.19}
}

@article{johnson1988generating,
  title={On generating all maximal independent sets},
  author={Johnson, David S. and Yannakakis, Mihalis and Papadimitriou, Christos H.},
  journal={Information Processing Letters},
  volume={27},
  number={3},
  pages={119--123},
  year={1988},
  publisher={Elsevier}
}

@inproceedings{schaefer1978complexity,
  title={The complexity of satisfiability problems},
  author={Schaefer, Thomas J.},
  booktitle={Proceedings of the tenth annual ACM symposium on Theory of computing},
  pages={216--226},
  year={1978}
}

@article{strozecki2019survey,
  title={Enumeration Complexity},
  author={Strozecki, Yann},
  journal={Bulletin of EATCS},
  volume={1},
  number={129},
  year={2019}
}

@article{tovey1984simplified,
  title={A simplified {NP}-complete satisfiability problem},
  author={Tovey, Craig A.},
  journal={Discrete applied mathematics},
  volume={8},
  number={1},
  pages={85--89},
  year={1984},
  publisher={Elsevier}
}

\end{document}